\documentclass[onecolumn,showpacs,preprintnumbers,amsmath,floatfix]{revtex4}
\usepackage{graphicx}
\usepackage{dcolumn}
\usepackage{bm}

\begin{document}

\title{Constraining low scale gravity with ultrahigh energy neutrinos}

\author{Shahid Hussain}
 \email{vacuum@ku.edu}
\author{Douglas W. McKay}%
 \email{mckay@kuark.phsx.ku.edu}
\affiliation{Department of Physics \& Astronomy 
University of Kansas, Lawrence, KS 66045.}

\date{\today}

\begin{abstract}
We show that LSG model predictions of event ratios are clearly distinguished 
from those of the SM. This is true in all models of ultrahigh energy (UHE)
 neutrino sources, in 
both $\nu_e:\nu_{\mu}:\nu_{\tau} = 1:1:1$ and $\nu_e:\nu_{\mu}:\nu_{\tau} = 1:2:0$
 scenarios for the 
flux incident on earth. In particular the ratios of upward $\mu$ events to upward 
shower events and the ratios of up events to down events are different 
by a factor of 2 to an order of magnitude in the comparisons between SM and LSG. 
$\nu_{\tau}$ rates are low but show high sensitivity to SM vs LSG interaction physics.
\end{abstract}
\maketitle

\section{Introduction}

There are a number of models predicting the existence of UHE (energies of a
PeV and higher) neutrinos. It is very
likely that UHE neutrinos exist as UHE cosmic rays have been observed and
most of the models that account for these cosmic rays also guarantee the existence
of UHE neutrinos.

Let us assume these neutrinos exist and we want to detect them. We know
nothing about their interactions at UHE energies and we do not know their
fluxes either. In this paper we differentiate, independent of the input
neutrino flux model, between two possibilities for UHE neutrino-nucleon
interactions. One is an extrapolation to of standard model (SM)
weak interaction cross sections to ultrahigh energies 
and the other is anomalous rise in the cross
sections at ultrahigh energies as predicted by low scale gravity (LSG) models
\cite{add,rs,aadd}.
Although we consider a particular low scale gravity scenario with large
extra dimensions, our results are applicable to any models which give
similar cross sections at UHE energies. 

For an analysis that does not depend on the normalization of the input flux,
we look at up-to-down ratios of the muon, tau, and shower events for an
icecube-like detector\cite{mcsh1,mcpj}. To do that, one needs to know
upward and downward neutrino flux at the detector site (south pole for icecube) 
for a given flux model. 
For input flux we consider the models SDSS\cite{sdss}, 
WB\cite{wb}(WB flux bound), Protheroe\cite{rjproth}, Mannheim (B)
\cite{manb} models.

The Earth is nearly opaque to UHE neutrinos (Earth is almost transparent to
neutrinos of energies below a 50TeV or so) and hence we solve
integro-differential equations for neutrino propagation through the Earth to
calculate the upward neutrino fluxes (neutrino fluxes coming through the
earth) at the detectors. In Section II we look at the cross sections
involved in neutrino-nucleon interactions, and propagation of the flux
through the earth. Section III\ has results and discussion for event rates,
and Section IV gives the summary.

\section{Neutrino propagation through the earth}

In LSG models, in addition to the SM neutral current (NC) and charged
current (CC) weak interactions, neutrinos interact with nucleons via
graviton exchange or they can form micro black holes. Graviton exchange
process is calculated in the eikonal (EK) approximation, and for black hole
formation (BH) one uses the geometric cross section. Here we are interested
in the detectable effects of these processes without going into theoretical
details of the processes. The EK process will initiate a particle shower. The
BH process also gives a particle shower via Hawking radiation, as the micro
black hole, unlike a macro black hole, has extremely short life time due to
its tiny mass. Presumably these particle showers will produce
radio and optical radiation which can be detected.

For UHE neutrinos, LSG models predict dominance of graviton exchange (EK)
and black hole formation (BH) over the extrapolated standard model
interactions. However, these models do not set the energy scale at which
this happens. Also, the number of extra dimensions is a free parameter. In
this paper we look at the interesting case of 1 TeV and 2 TeV scale LSG
models as this scale is close to the weak scale and, also as we will see
later, UHE neutrinos can be useful to give us signals of LSG only around
this scale. Reference\cite{mcsh1} has the cross sections and interaction lengths for 
neutrino-nucleon interactions in SM
and LSG. 
To calculate the upward neutrino flux at the detector we
need to propagate not only three neutrino flavors but also the $\tau$'s produced
in CC interactions of $\nu{\tau}$'s\cite{mcsh1}. These taus will decay and produce more
neutrinos of all flavors hence we cannot ignore them. We do not need to
propagate electrons, muons, and hadrons produced by neutrinos during their
propagation, as excluding them does not have any significant effect on the
neutrino flux propagation. The numerical solution of the coupled
integro-differential equations of propagation in SM and LSG, for three
neutrino flavors and the taus, gives us the final upward neutrino flux at
the detector site. This flux will be different in SM and LSG\ due to
different interactions involved in the neutrino propagation. The over all
effect of LSG is to give a stronger suppression of the upward flux as
compared to the SM case.

\begin{table}
\caption{\label{tab:table1}Up and down events $yr^{-1}$ (in the flux models SD\cite{sdss} 
(or SDSS), 
WB\cite{wb}, PR\cite{rjproth}, and MB\cite{manb}) 
for detector threshold $E=0.5PeV$ 
in the scenario $\nu_e:\nu_{\mu}:\nu_{\tau} = 1:1:1$; results are shown for 
the standard model (SM), and low scale gravity with 6 extra dimensions and mass scales 
1TeV (G1) and 2TeV (G2) (for details see Ref.\cite{mcsh1}). 
All upward events are integrated over nadir angle $\theta \leq 84^{0}$; down showers 
are integrated over angle  but muons and taus are not.}
\begin{tabular}{|c|c|c|c|}
\hline
& showers ($\frac{up}{down}$) & muons ($\frac{up}{down}$) & taus ($\frac{up}{%
down}$) \\ \hline
$
\begin{array}{l}
\\ 
WB\smallskip \\ 
SD \smallskip \\ 
MB \smallskip \\ 
PR\smallskip 
\end{array}
$ & $
\begin{array}{lll}
SM & G2 & G1 \\ 
\frac{2.6}{11}\smallskip  & \frac{2.7}{24} & \frac{3.1}{201} \\ 
\frac{163}{622}\smallskip  & \frac{167}{748} & \frac{202}{4725} \\ 
\frac{3.0}{30}\smallskip  & \frac{3.1}{195} & \frac{2.2}{1898} \\ 
\frac{32}{182}\smallskip  & \frac{33}{534} & \frac{34}{5284}
\end{array}
$ & $
\begin{array}{lll}
SM & G2 & G1 \\ 
\frac{3.0}{3.3}\smallskip  & \frac{2.7}{3.3} & \frac{1.1}{3.3} \\ 
\frac{176}{142}\smallskip  & \frac{165}{142} & \frac{74}{142} \\ 
\frac{6.4}{18}\smallskip  & \frac{3.6}{18} & \frac{0.66}{18} \\ 
\frac{50}{73}\smallskip  & \frac{38}{73} & \frac{12}{73}
\end{array}
$ & $
\begin{array}{lll}
SM & G2 & G1 \\ 
\frac{0.11}{0.13}\smallskip  & \frac{.074}{0.13} & \frac{0.0086}{0.13} \\ 
\frac{4.7}{4.6}\smallskip  & \frac{4.0}{4.6} & \frac{0.54}{4.6} \\ 
\frac{0.46}{0.86}\smallskip  & \frac{0.20}{0.86} & \frac{0.01}{0.86} \\ 
\frac{2.4}{3.5} \smallskip & \frac{1.5}{3.5} & \frac{0.13}{3.5}
\end{array}
$ \\ \hline
\end{tabular}
\hspace{0pt} 
\end{table}

\section{Event rates}

With the propagated flux in hand, we calculate the upward and down shower,
muon, and tau event rates\cite{mcsh1} due to different neutrino interactions in SM and
LSG. An ICECUBE-like detector is capable of detecting muons by their tracks,
taus by some tagged tau events as explained below, and showers by the
optical radiation they produce.

There are two sources of muon events: (i) muons from $\nu{\mu}$-nucleon CC\
interaction (ii) muons from decays of the taus produced in a $\nu{\tau}$-nucleon CC
interaction. As LSG does not introduce any interaction analogous to CC
interaction in SM, one expects the muon events will be the same in SM and
LSG provided the neutrino flux is the same in the two. This means the down
muon events will be the same in SM and LSG but upward muon events will be
smaller in LSG as the upward neutrino flux is smaller in LSG as compared to
the one in SM.

The only source of taus is $\nu{\tau}$-nucleon CC interaction. As explained below,
tagging taus is harder than tagging muons, as one cannot tag taus by
detecting their tracks alone. Fortunately, there are other
ways to tag tau events by detecting\cite{mcsh1}: (i) a track and a particle shower
at the end of the track.(ii) a shower, a track,
and then a shower again.(iii) a tau track and the muon track after
muonic tau decay. However,
it turns out that all the above three event types for tagging taus are only
a few in number for any neutrino flux model. For the same reason as
mentioned above for muons, the tagged tau down events will be the same in SM
and LSG but the upward events will be larger in SM as compared to LSG.

Shower events are detected by the radiation coming from the particle
showers. These particle showers are produced in all neutrino-nucleon
interactions in SM (NC+CC) and LSG (EK+BH). However, for a given neutrino
flux, we expect to see more showers in LSG as compared to SM; This is
because for LSG we have, in addition to CC\ and NC interactions, much bigger
EK and BH cross sections.

In table I we show up-to-down event rates for showers, muons, and taus in SM
and LSG for different input flux models. Here we see the shower muon rates
are significant for all the flux models and both in MS and LSG. One the
other hand the tau rates are very small as we anticipated. In table II, to
better differentiate between SM and LSG, we show some ratios of the event
rate ratios in the flavor scenarios $\nu_e:\nu_{\mu}:\nu_{\tau}::1:2:0$ and
$\nu_e:\nu_{\mu}:\nu_{\tau}::1:1:1$ for the input flux at the Earth. Here we see the ratio
of the showers down to muons up is more than an order of magnitude larger in
LSG, with 1 TeV mass scale, as compared to SM. This is true in both of the
flavor scenarios. For the 2 TeV scale gravity, the difference between SM and
LSG are not as large. As we see in table II, the up-to-down tagged tau event
ratios will also be very useful in constraining low scale gravity.

\begin{table}
\caption{\label{tab:table2}Ratios of the ratios (detector threshold $E=0.5PeV$); here 
$RR1$=$\frac{showers\text{ }down}{muons\text{ }up}$ and 
$RR2$=$\frac{taus\text{ }down}{taus\text{ }up}$. Results are shown fo two different flavor 
scenarios.}
\begin{tabular}{|c|c|c|c|c|}
\hline

&$\frac{RR1_{G2}}{RR1_{SM}}$ & $\frac{RR1_{G1}}{RR1_{SM}}$ & 
$\frac{RR2_{G2}}{RR2_{SM}}$ & $%
\frac{RR2_{G1}}{RR2_{SM}}$ \\ \hline
$
\begin{array}{l}
\\
WB\smallskip \\ 
SD \smallskip \\ 
MB \smallskip \\ 
PR\smallskip 
\end{array}
$ & $
\begin{array}{ll}
1:2:0 & 1:1:1 \\ 
\frac{4.5}{1.7}\text{=2.7} \smallskip & \frac{8.8}{3.6}\text{=2.4} \\ 
\frac{2.1}{1.6}\text{= 1.3} \smallskip & \frac{4.5}{3.5}\text{=1.3} \\ 
\frac{30}{2.1}\text{= 14} \smallskip & \frac{54}{4.7}\text{=11} \\ 
\frac{7.4}{1.6}\text{=4.7} \smallskip & \frac{14}{3.6}\text{=3.9}
\end{array}
$ & $
\begin{array}{ll}
1:2:0 & 1:1:1 \\ 
\frac{93}{1.7}\text{=56} \smallskip & \frac{183}{3.6}\text{=50} \\ 
\frac{34}{1.6}\text{= 21} \smallskip & \frac{64}{3.5}\text{=18} \\ 
\frac{1571}{2.1}\text{= 735}\smallskip  & \frac{2875}{4.7}\text{=610} \\ 
\frac{249}{1.6}\text{=157}\smallskip  & \frac{440}{3.6}\text{=121}
\end{array}
$ & $
\begin{array}{l}
\\
1.5 \smallskip \\ 
 1.2 \smallskip \\ 
2.3 \smallskip \\ 
1.6\smallskip 
\end{array}
$ & $
\begin{array}{l}
 \\ 
13\smallskip  \\ 
 8.7 \smallskip \\ 
 46 \smallskip \\ 
18\smallskip 
\end{array}
$ \\ \hline
\end{tabular}
\vspace*{0pt}
\end{table}

\section{Summary}

The following comments are true for all the neutrino flux models we have
considered (see Ref.\cite{mcsh1}): (i) Even at thresholds as large as 0.5PeV, 
the showers and muon
event rates are large enough to give a significant signal in an ICECUBE-like
detector. (ii) Although LSG (1-2TeV scale) sets in around 2-20PeV neutrino
energies, it is possible (due to neutrino feed down effect) to distinguish
between SM and LSG (1-2TeV) even at lower energy thresholds. Even at energy
thresholds as low as 0.5PeV the event ratios of showers down to muons up are
significantly large in LSG (1-2TeV) as compared to SM. This is true for both
flavor scenarios ( $\nu_e:\nu_{\mu}:\nu_{\tau}::1:2:0$ and 
$\nu_e:\nu_{\mu}:\nu_{\tau}::1:1:1)$. (iii)
Tagged tau neutrino events, although very small, are useful in
differentiating between SM and LSG (1TeV). The event ratios of taus down to
taus up are significantly large in LSG (1TeV) as compared to SM. The taus
are not a very useful tool in differentiating between SM and LSG with scale
of 2TeV and higher. (iv) To differentiate between SM and LSG with scales
higher than 2TeV, one needs to look at events with thresholds higher than
10PeV which makes it very hard as the neutrino fluxes fall very rapidly at
higher energies.

\section*{Acknowledgments}
Shahid thanks the LLWI (2004) organizers, the University of Kansas 
Student Travel Fund, and the high energy group for support. This work 
was supported in part by the U.S. Department of Energy under 
Grant No. DE-FG03-98ER41079.

\end{document}